\documentstyle[11pt,paspconf,epsf]{article}

\begin{document}
\title{Mass Distribution in Compact Groups }
\author{J. Perea,  A. del Olmo, L. Verdes-Montenegro}
\affil{Instituto de Astrof\'{\i}sica de Andaluc\'{\i}a (CSIC), Granada, Spain}
\author{M. S. Yun}
\affil{National Radio Astronomy Observatory,\altaffilmark{1}, Socorro,
NM, USA}
\author{W.K. Huchtmeier}
\affil{Max-Planck-Institut f\"{u}r Radioastronomie, Bonn, D-53121, Germany}
\author{B. A. Williams}
\affil{University of Delaware, Newark, Delaware, USA}
\altaffiltext{2}{The National Radio Astronomy Observatory is a
facility of the National Science Foundation operated under cooperative
agreement by Associated Universities, Inc.}

\begin{abstract}
New redshift surveys of galaxies in the field of compact groups have
discovered a population of faint galaxies which act as satellites
orbiting in the potential well of the bright group. Here we analyze
the mass distribution of the groups by comparing the mass derived from
the bright members and the mass obtained from the satellite
galaxies. Our analysis indicates the presence of a dark halo around
the main group with a mass roughly four times that measured for the
dominant galaxies of the compact group.

We found that heavier halos are ruled out by the observations when
comparing the distribution of positions and redshifts of the satellite
galaxies with the distribution of satellites surrounding isolated spiral
galaxies.  The results agree with a picture where compact groups may
form a stable system with galaxies moving in a common dark halo.

\end{abstract}

\section{Introduction}
As it is known, the existence of compact groups poses an interesting
dynamical problem. Early simulations such as those done by Barnes
(1989) suggested that the time scale for merging was very short.  This
result has lead some authors to question the reality of the groups
or to describe them as transient configurations in larger systems
(Mamon, 1986; Hernsquist et al. 1995; Diaferio et al. 1994). However
recent simulations showed that there exist stable enough systems with
a common dark halo around all the galaxies (Athanassoula et al. 1997).

New observations indicate the existence of fainter galaxies at the
same redshift of the group but at larger distances (e.g. de Carvalho
et al. 1997, Barton et al., 1998, Zabludoff \& Mulchaey, 1998). In
this work, we assume that the small galaxies are orbiting as test
particles in the potential well of the bright group. For such
configuration we can obtain two mass estimations, one from the
galaxies forming the compact group and the other from the positions
and velocities of the satellites We expect the last to be larger, if a
dark halo exist, since we are sampling a larger scale.

\section{The Sample} 

We analyze 13 compact groups from the Hickson's catalog (1982).
Redshifts and positions for satellites were obtained for groups
HCG~16, 23, 42, 62, 63, 67, 86, 87, 90, and 97 from the work by De
Carvalho et al (1997). In addition we included HCG~96, 92 and 37,
where the redshifts were obtained from our VLA HI data (HCG~96 and 92)
and from optical data obtained by Alfosc\footnote{The data presented here
have been taken using ALFOSC, which is owned by the Instituto de
Astrof\'{\i}sica de Andalucia (IAA) and operated at the Nordic Optical
Telescope under agreement between IAA and the NBIfA of the
Astronomical Observatory of Copenhagen.} at the Nordic Optical
Telescope (HCG~37 and 96). We also included the objects detected by
Peterson \& Shostak (1980) in HCG~92.

\section{Mass estimation and dark halos}

To measure the mass of each group, we evaluated the two versions of the
projected and virial mass estimators. For the compact group the self
graviting estimators of the virial ($M_v(CG)$) and projected
($M_p(CG)$) mass were calculated after Heisler et al. (1985) and Perea
et al. (1990). For the satellites, we applied the mass estimators
($M_v(Orb)$ and $M_p(Orb)$) for test particles around a point mass as
described by Bahcall \& Tremaine (1981).

\begin{figure}[htb]
\plotfiddle{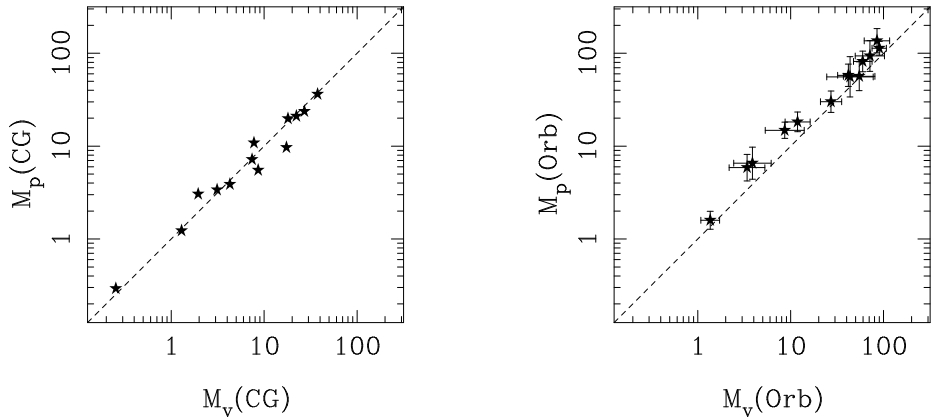}{5.2cm}{0}{100}{100}{-220}{-25}
\caption{Comparison between the Virial and Projected Mass. The
error bars were obtained by bootstrap.}
\label{fig-1}
\end{figure}

In Fig.\ \ref{fig-1} we compare the virial and projected mass
estimators. In the left panel we show the results for the compact
groups (CG) and in the right one the estimations from the satellite
galaxies (Orb), where the mass is expressed in units of 10$^{12}$M$_{\odot}$
with H$_{0}=50$ km~s$^{-1}$Mpc$^{-1}$.  As can be seen, both
estimators are equivalent for CGs but there is a discrepancy for the
orbiting satellite galaxies where we obtain $M_{p}(Orb)\approx
1.37\times M_{v}(Orb)$.

\begin{figure}[htb]
\plotfiddle{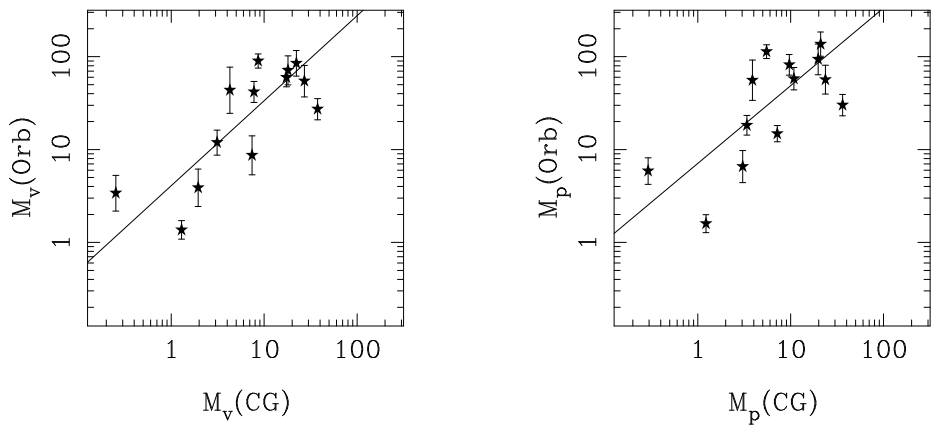}{5.2cm}{0}{100}{100}{-220}{-25}
\caption{Comparison between the orbital and CG masses}
\label{fig-2}
\end{figure}

The orbital mass is a linear function of the central mass as 
expected when the compact group dominates the dynamics. As can be seen
from Fig.\ \ref{fig-2}, the results are consistent with an almost
constant fraction $M(Orb)/M(CG)$. Observationally we found the
following linear fit,
$$M_{v}(Orb) \approx (4.1\pm 0.7)\times M_{v}(CG)^{0.92\pm 0.07}$$
with a Student's t=4.0 (probability $>99\%$) for the correlation.
This result indicates the existence of a dark halo with a larger
extension, concentrated on the position of the compact group, and
with a mass about four times the one obtained from the
dominant galaxies of the compact group.

\subsection{The extent of the halos}
\begin{figure}
\plotfiddle{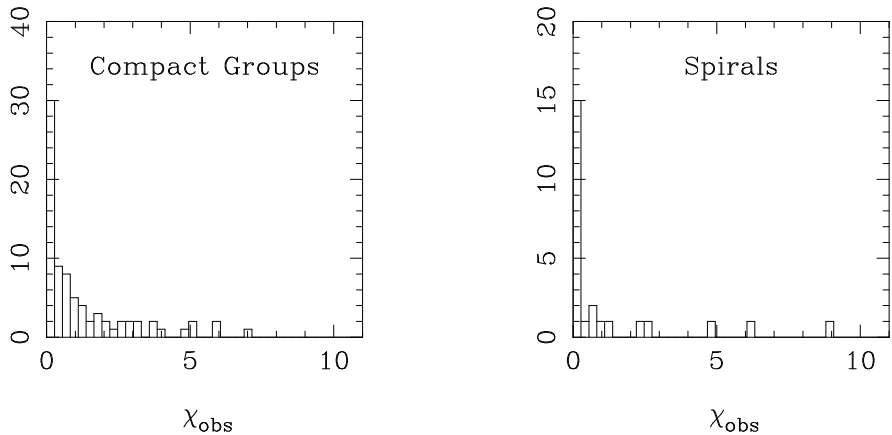}{5.6cm}{0}{100}{100}{-220}{-20}
\caption{$\chi_{obs}$ for Compact Groups and Spirals}
\label{fig-3}
\end{figure}

It is possible to know more about the extent of the dark halo if we
combine all the information provided by the satellites. In doing
so we followed the formalism derived by Van Moorsel (1982) for binary
galaxies and by Erickson et al. (1999, EGH) for satellites of spiral
galaxies.  We analyze the distribution of orbital masses as defined by
$M_{\chi}=v_{z}^{2}r_{p}/G$ using the satellites in each compact group
where $v_{z}$ is the radial velocity of the satellite with respect to
the central group and $r_{p}$ is the projected separation. The orbital
mass should be corrected by a factor $\chi$, accounting for all the
projections, to obtain the real mass of each group (see EGH, for
details) . For that we use the observational quantity
$\chi_{obs}=v_{z}^{2}r_{p}/M_{g}$, related to $\chi$ through
$\chi_{obs}=(M_{\chi }/M_{g})\chi$, where $M_{g}$ is the mass of the
central system ($M_{v}(CG)$ or $M_{p}(CG)$). In this way the analysis
of the mass distribution is reduced to the study the distribution
of $\chi_{obs}$ for all the satellite galaxies in all groups (see
left panel of Fig. \ref{fig-3}). The quantity $\chi_{obs}$ measures
the extension of the dark halo.

The $\chi_{obs}$ distribution for the satellite galaxies in CGs are
similar to those observed for the satellites of isolated spirals by
EGH (right panel in Fig.\ \ref{fig-3}) and in particular their models
5 \& 6 apply here and indicate that the values of $\chi_{obs}$ can be
explained only if no heavy halos are present.

\acknowledgements 
The authors acknowledge interesting discussions with J. Sulentic,
E. Athanassoula and A. Bosma. JP, AO, and LV--M, are supported
by DGICYT Grant PB96-0921 and Junta de Andaluc\'{\i}a (Spain).

\end{document}